\documentclass[journal=ancham, manuscript=article, layout=twocolumn]{achemso}

\usepackage[utf8]{inputenc}
\usepackage[T1]{fontenc}
\usepackage[british]{babel}

\usepackage{amsmath, graphicx, subcaption}
\usepackage[locale=UK, retain-unity-mantissa=false]{siunitx}

\title{Validation of secondary fluorescence excitation in quantitative X-ray fluorescence analysis of thin alloy films}
\author{André Wählisch}
\email{andre.waehlisch@ptb.de}
\author{Cornelia Streeck}
\author{Philipp Hönicke}
\author{Burkhard Beckhoff}
\affiliation[PTB]{Physikalisch-Technische Bundesanstalt\\Abbestr. 2-12\\10587 Berlin, Germany}

\begin{document}

\maketitle

\begin{abstract}
X-ray fluorescence (XRF) analysis is a widely applied technique for the quantitative analysis of thin films up to the \si{\um}~scale because of its non-destructive nature and because it is easily automated. When low uncertainties of the analytical results in the few percent range are required, the non-linear secondary fluorescence effect in multi-elemental samples may complicate an otherwise straightforward quantification, since it can easily exceed a relative contribution of \SI{20}{\percent}. The conventional solution, to rely on good performing reference samples, is hindered by their low availability, especially for thin film applications. To address this challenge, we demonstrate a flexible production method of multilayered, alloyed thin films with significant secondary fluorescence contributions. We use reference-free XRF analysis to validate the reliability of the physical model for secondary fluorescence, which includes a thorough uncertainty estimation. The investigated specimens are being qualified as calibration samples for XRF or other quantitative analyses.
\end{abstract}

\section{Introduction}

Thin films with thicknesses from a few \si{\nm} to a few \SI{10}{\um} are used in many technical applications because of their useful electromagnetic or mechanical properties. Use cases are highly specialized: Automotive paints~\cite{Krimi2016}, consisting of multiple layers of several \si{\um} each, super-alloys for micro-electro-mechanical systems~\cite{Sim2017}, various integrated circuits~\cite{Zang2015}, and energy storage in thin film batteries~\cite{Li2017}, to list just a small selection. Accordingly, the analysis of these materials is crucial for process development and production control in a diverse spectrum of applications, where demands with respect to accuracy are steadily growing. This requires flexible and reliable analytical tools. While there is a plethora of different analytical techniques available~\cite{Abou-Ras2011,Abou-Ras2015}, the technique of choice depends both on the given sample and key parameters of interest. Photon induced XRF analysis is a non-destructive, non-preparative analytical technique, which can be applied to a wide range of different materials. It is employed by industrial and scientific laboratories using various excitation sources, such as conventional X-ray tubes, radioactive materials or synchrotron radiation beamlines. XRF can be used to identify chemical elements in a sample and to quantify the elemental mass deposition (mass per unit area) of individual layers in a thin film sample~\cite{Kolbe2005a,Vrielink2012,lessThanInfinite}.

Quantitative results in XRF analysis are usually achieved by utilizing appropriate reference materials, which are measured under the same conditions as the sample of interest. This can compensate for missing knowledge on experimental or instrumental parameters and often allows for accurate and precise quantitative results if the reference sample is well-characterized with low uncertainties. In principle, when using several appropriate reference samples the accuracy of the results can be improved even further~\cite{handbook}. A general constraint of this approach is given by the need for suitable reference samples, since their chemical and spatial composition need to match the sample of interest rather closely. The supply of such appropriate reference samples may not be growing as fast as their demand in rapidly advancing technological fields. For multilayer systems consisting of pure elements, examples of reference standards do exist~\cite{Hoffmann2003,nygard,Sakurai2019}. However, in the case of multilayers consisting of alloyed material, appropriate reference materials are rare~\cite{lessThanInfinite}. Reference-free XRF~\cite{Beckhoff2008b}, as applied by the Physikalisch-Technische Bundesanstalt~(PTB), relies on calibrated instrumentation instead of reference materials and can serve as an alternative approach. In practice~\cite{Scholze2009,Krumrey2006,Gottwald2006} this is realized with calibrated photodiodes and calibrated silicon drift detectors~(SDD) as well as good knowledge on the atomic fundamental parameters. Since this method does not require reference samples to derive quantitative results, it can also be used to qualify samples as calibration samples for XRF analysis or other analytical methods~\cite{Hoenicke2018}.

In multi-elemental materials the fluorescence radiation of one element may not only absorb but also enhance the fluorescence emission of another. This secondary fluorescence effect can easily contribute more than \SI{20}{\percent} to the total fluorescence intensity of an individual element, depending on the elemental distribution, the spatial dimensions of the system, the angle of incidence and of detection as well as the energy of the excitation radiation. In this case, the usage of a proper physical model of this matrix effect, describing the interactions of X-ray radiation with matter in the sample, is crucial for achieving accurate quantitative analytical results. Utilizing reference-free XRF experiments with tunable synchrotron radiation we validate the commonly employed secondary fluorescence model by quantitatively analysing samples exhibiting considerable secondary fluorescence contributions of about \SIrange{10}{30}{\percent}. The quantitative results are examined in respect to their physical consistency under variation of the experimental conditions and in comparison to prior knowledge about the samples: Tuning the energy of the monochromatic excitation radiation or altering the angle of incidence changes the contribution of secondary fluorescence radiation significantly. We verify that the physical model adequately captures these changes. Firstly, we analyse commercially available thin foils of pure materials. When putting them on top of each other, they create a simple multilayer inducing secondary fluorescence radiation. Since the foils can be measured separately, allowing for an independent quantitative characterization, this provides a direct validation approach. Furthermore, we analyse systematically deposited thin films on a wafer. Particular attention is given to the design and characterization of alloy materials, in view of the lack of thin alloy reference samples.

\section{Quantitative XRF analysis of layered and alloyed material}

Quantitative XRF analysis of multilayer systems aims to determine elemental mass depositions of individual layers, and is based on the evaluation of X-ray fluorescence intensities~\cite{handbook}. The analytical equation to calculate the intensity of an individual fluorescence line for a given sample composition is given by \citeauthor{sherman}~\cite{sherman}. Approximations exist for the limiting cases of bulk and thin samples~\cite{VanGrieken2001}, but equations are also known for the interesting case of multilayers with intermediate thicknesses~\cite{Mantler1984,Mantler1986,boer,boerkorrektur}. These equations  are based on the fundamental parameter~(FP) approach. It requires good knowledge about the atomic fundamental parameters describing the probabilities involved in the fluorescence process, such as photo-absorption coefficients, fluorescence yields, and transition probabilities. When neglecting tertiary~\cite{Shiraiwa1966} and higher excitation effects, the total intensity~$I_i$ of the fluorescence line~$i$ due to the excitation with monochromatic radiation is given by
\begin{equation}
I_i=P_i+\sum_{j}S_{ij}\,,\label{eq: intensity}
\end{equation}
where $P_i$ is the intensity of primary fluorescence, i.e., excitation due to the external radiation. The sum is evaluated for all energetically possible secondary enhancements~$S_{ij}$ due to the excitation by fluorescence line~$j$. Definitions of primary excitation~$P_i$ and secondary excitation~$S_{ij}$ are given, e.g., by \citeauthor{boer}~\cite{boer}. For multilayer samples, the equations need to take into account the absorption effects of all layers, and all energetically possible interlayer and intralayer excitations. This results in a highly non-linear system of equations with respect to the elemental mass depositions, where the fluorescence intensity of one element can depend on the fluorescence intensities of multiple other elements. Thus, there exists no known solution to directly determine the elemental mass deposition from the total fluorescence intensity for multi-elemental multilayers. However, if all experimental parameters and atomic FPs in equation~\eqref{eq: intensity} are known, the mass deposition can be estimated with a non-linear optimization algorithm~\cite{boerkorrektur,Bos1998}.

Furthermore, the total uncertainty of the mass deposition needs to be determined. Here, the uncertainties of all experimental parameters and FPs have to be taken into account. The propagation of uncertainty is not trivial for quantitative XRF analysis of stratified materials, where the relevant equation~\eqref{eq: intensity} cannot directly be solved for the mass deposition. Therefore, we use a Monte-Carlo~(MC) based approach, which is defined in the supplement~1 to the `Guide to the expression of uncertainty in measurement'~\cite{GUMsup}. The MC approach assigns a probability density function~(PDF) to each experimental parameter and FP. The PDF describes the likelihood of an input parameter value~$\mu$ to be in a specific range, as defined by the uncertainty~$\sigma$ of the parameter. We use a Gaussian input PDF with mean~$\mu$, determined by the actual value of the individual parameter. The variance~$\sigma^2$ of the Gaussian is derived from the parameter uncertainty~$\sigma$. While PTB is able to experimentally determine selected FPs and their uncertainties~\cite{Menesguen2015,Kolbe2015}, not all FPs involved in the current work have been experimentally determined. In this case, values are taken from the \emph{xraylib} database~\cite{xraylib}, and upper bounds on the uncertainties are derived based on estimations by \citeauthor{krause}~\cite{krause}. All input PDFs are simultaneously propagated through the theoretical model to obtain an estimated output PDF for the mass deposition. This output PDF allows for a straightforward statistical evaluation, e.g., the numerical calculation of mean and standard deviation, which are used as estimates for the mass deposition results and their uncertainties, respectively.

\section{Experimental design}

\begin{figure}
\centering
\includegraphics[keepaspectratio, width=\linewidth]{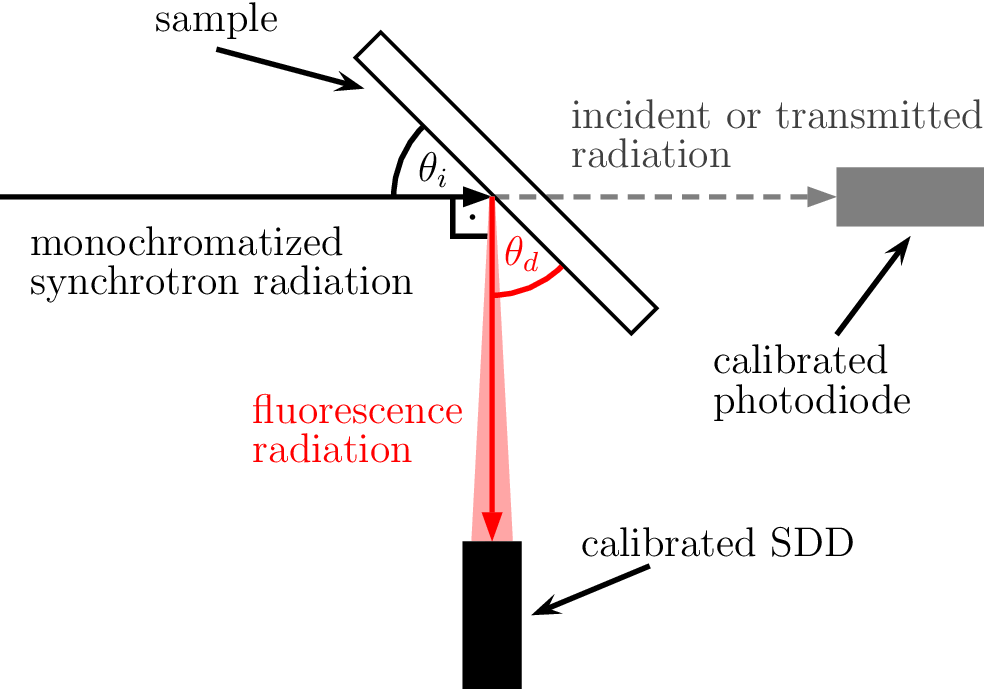}
\caption{Experimental design inside the UHV chamber for X-ray spectrometry. The intensity of the incident radiation can be monitored by moving the sample.}
\label{fig: expsetup}
\end{figure}

Reference-free XRF measurements were carried out at the four-crystal monochromator beamline~\cite{Krumrey1998} in the PTB laboratory and at the \emph{BAMline}~\cite{Riesemeier2005}, both located at the electron storage ring for synchrotron radiation BESSY~II. Transmission measurements were conducted at the \emph{BAMline}. Figure~\ref{fig: expsetup} shows a schematic view of the experimental design. A calibrated SDD~\cite{Scholze2009,Krumrey2006} with characterized detector response function and well-known spectral efficiency is used to collect the fluorescence radiation originating from the samples. The effective solid angle of detection is determined from the knowledge of the collimating geometry of the SDD~\cite{Beckhoff2007a}. These experimental parameters are used to quantitatively determine the photon flux of the fluorescence radiation with a spectrum deconvolution approach~\cite{Kolbe2010}. A calibrated photodiode~\cite{Gottwald2006} is utilized to monitor the incident photon flux of the monochromatized synchrotron radiation and the photon flux of the transmitted radiation. Detectors and samples are placed inside an ultra-high vacuum~(UHV) chamber~\cite{Lubeck2013}, where the pressure is kept below $\SI{1E-7}{\milli\bar}$ during measurements, to minimize the attenuation of X-ray radiation by air. The angles of incidence and detection were set to~$\theta_i=\theta_d=\SI{45}{\degree}$ for all but the angular dependent measurements.

\section{Secondary fluorescence of an artificially assembled multilayer}

For a demonstrative approach to the validation of the secondary fluorescence model we studied two commercially available thin foils. These consist of copper and titanium respectively, and when one is put on top of the other, create an artificial multilayer sample. Copper may cause significant titanium secondary fluorescence in this multilayered system. A quantitative XRF analysis performed under such conditions will only give correct results if the secondary fluorescence effect is properly taken into account. Conversely, if the mass depositions of the separate foils are known, this measurement can be used to validate the secondary fluorescence model. The foils have a high chemical purity (better than \SI{99.97}{\percent}), and thicknesses of nominally $d_{\text{Cu}}=\SI{10}{\um}$ and $d_{\text{Ti}}=\SI{4}{\um}$. Mass deposition $\rho{}d$ and homogeneity of each separate foil are rapidly determined by transmission measurements utilizing monochromatic X-ray radiation and experimentally determined FPs~\cite{Menesguen2015}. After that, the titanium foil is put on top of the copper foil, and the stack is analysed with reference-free XRF experiments. The secondary fluorescence effect must be taken into account only in this case. The transmission measurement of the titanium foil was performed with an excitation energy of $E_0=\SI{8.4}{\keV}$, the measurement of the copper foil at $E_0=\SI{13.5}{\keV}$. These values were chosen to reduce the uncertainty of the transmission measurement~\cite{nordfors,chantlerNordfors}. The reference-free XRF measurement was carried out with an excitation energy of $E_0=\SI{15}{\keV}$. The relative contribution of secondary fluorescence radiation to the total K$\alpha$ fluorescence intensity of titanium is approximately \SI{27}{\percent} under the given experimental conditions.

Transmission and XRF measurements of the thin foil samples were not necessarily conducted on identical sample positions. To ensure the comparability of the results the homogeneity was ascertained by varying the lateral measurement position on the samples with a step size of~\SI{250}{\um}. A relative standard deviation for the transmitted intensity of $<\SI{0.7}{\percent}$ was determined and deemed satisfactory for all further results. For the reference-free XRF measurements only the titanium mass deposition is determined, and copper assumed known from the transmission measurement. This approach is necessary because the copper at the bottom of the layer stack is close to the saturation thickness~\cite{VanGrieken2001} of the incident X-ray radiation. The mass deposition~$\rho{}d$ of each individual foil can be converted into a thickness~$d$ by assuming only negligible deviations of the density~$\rho$ from the bulk density~$\rho_b$, i.e., $\rho=\rho_b$. The resulting titanium thickness from the transmission measurement is $d_{\text{Ti}}=\SI{3.8(2)}{\um}$ and from the XRF measurement $d_{\text{Ti}}=\SI{3.6(3)}{\um}$. These results are consistent within the calculated uncertainties, and in that sense validate the secondary fluorescence model. This validation methodology is readily extensible, since a multitude of combinations of thin foils are realizable. The commercial foils are affordable and chemically pure, which provides many possibilities for such an approach. The foils can also be analysed from both sides with XRF measurements, by just reversing the layer order. This changes the absorption paths, and the secondary fluorescence contributions in the sample substantially. If the samples are sufficiently homogeneous, which is easily verified, this must produce equal mass deposition results, providing a flexible and thorough validation approach.

\section{Secondary fluorescence of alloyed multilayers}

To investigate alloyed materials we rely on thin multilayers produced by ion beam sputter deposition~(IBSD). This method allows for the production of smooth surfaces with little defects~\cite{Gawlitza2006} for many chemical elements. Samples were created by the \emph{Fraunhofer-Institut für Werkstoff- und Strahltechnik} (IWS) in Dresden. In the IBSD system, different sputter targets can be placed on a rotating table, and the desired target material is selected by rotation. Thus, multiple elements can be deposited successively without breaking the required ultra-high vacuum, i.e., under nearly unaltered conditions. Layers of single elements and multilayers were created by utilizing an aperture to control the location of deposition: Three transition metals, titanium, chromium and copper, were deposited onto a silicon wafer (\SI{76.2}{\mm} diameter) on multiple patches and in various combinations. The patches have an area of approximately~$5\times\SI{15}{\mm\squared}$ each and were capped with a \SI{5}{\nm} carbon layer for protection against oxidation. This capping layer is transparent for the hard X-ray radiation involved in all presented experiments and is only mentioned for the sake of completeness. In a sample consisting of all three metals, copper may induce secondary K fluorescence for titanium and chromium, and chromium may induce secondary K fluorescence for titanium.

\begin{figure}
\centering
\includegraphics[keepaspectratio]{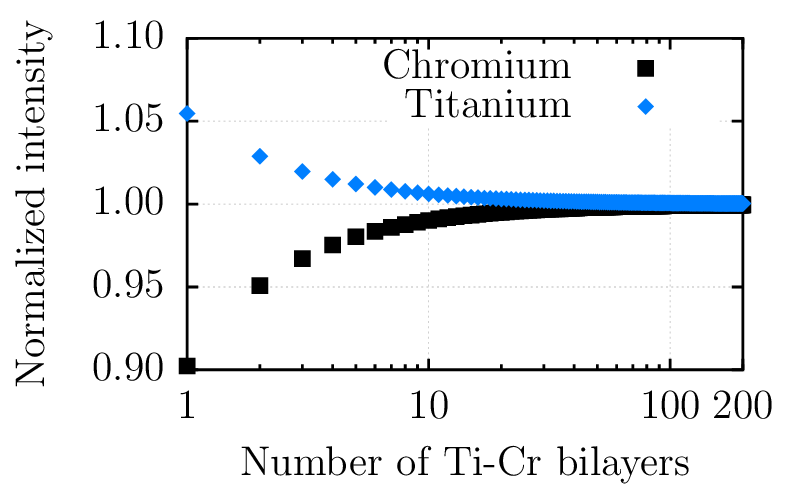}
\caption{Calculated titanium and chromium K$\alpha$ fluorescence intensities of multilayer samples with an increasing number of sublayers. The intensities are normalized to the respective fluorescence intensities of a single homogeneous alloy layer with equal mass depositions. The angle of incidence and detection were set to $\theta_i=\theta_d=\SI{45}{\degree}$. The monochromatic excitation energy was set to \SI{10}{\keV}.}
\label{fig: numberoflayers}
\end{figure}

As an appropriate surrogate for an alloy, a succession of chromium and titanium layers were produced by alternating the chromium and titanium deposition with short deposition duration. Bilayers created in this way have thicknesses of a few nanometres each, and the resulting sequence of \num{200}~titanium-chromium bilayers has a total thickness in the order of about~\SI{1}{\um}. The fluorescence radiation emitted by these \num{200}~bilayers is practically identical to that emitted by a single homogeneous alloy layer of equal mass depositions. This assumes as a prerequisite that the total thickness is well below the saturation thickness~\cite{VanGrieken2001} of the involved X-ray radiation, which is true for all measurements in this work. For demonstration, figure~\ref{fig: numberoflayers} shows the calculated titanium and chromium fluorescence intensities for an increasing number of bilayers in a multilayer, for which the mass depositions are fixed to \SI{0.2}{\milli\g\per\centi\metre\squared} for titanium and \SI{0.4}{\milli\g\per\centi\metre\squared} for chromium. The intensities are normalized to the intensity emitted by a single homogeneous alloy layer with identical mass depositions. If both sample systems are equivalent, this ratio ought to converge to unity. For \num{200}~bilayers, the ratio is calculated to be about \num{1.0003} for titanium and \num{0.9995} for chromium. Thus, the present samples can be considered uniform alloys for all intents and purposes of quantitative XRF analysis. In principle, this method allows for the production of well-defined and nearly arbitrary alloyed samples.

\begin{figure}
\centering
\subcaptionbox{Separate Cu layer.\label{subfig: drawing_cu}}[0.485\linewidth]{%
\includegraphics[keepaspectratio]{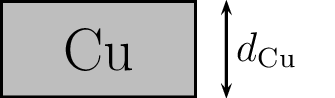}
}
\hfill
\subcaptionbox{Separate Ti-Cr alloy.\label{subfig: drawing_ticr}}[0.485\linewidth]{%
\includegraphics[keepaspectratio]{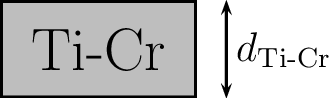}
}
\hfill
\subcaptionbox{Copper layer below Ti-Cr alloy.\label{subfig: drawing_ticrcu}}[0.485\linewidth]{%
\includegraphics[keepaspectratio]{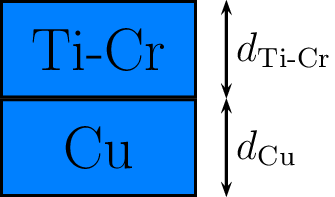}
}
\hfill
\subcaptionbox{Copper layer on top of Ti-Cr alloy.\label{subfig: drawing_cuticr}}[0.485\linewidth]{%
\includegraphics[keepaspectratio]{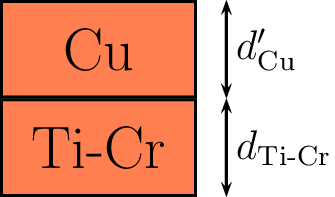}
}
\caption{Composition and layer sequence of the different types of analysed samples.}
\label{fig: schematics}
\end{figure}

The analysed samples consist of a pure copper layer (figure \ref{subfig: drawing_cu}), a sample of the \num{200} titanium-chromium bilayers separately (figure \ref{subfig: drawing_ticr}), a multilayer sample with the combination of both, the copper layer on the bottom (figure \ref{subfig: drawing_ticrcu}), and a multilayer sample with the copper layer on the top (figure \ref{subfig: drawing_cuticr}). Both separate copper layer, and the copper at the bottom of one of the multilayers (\ref{subfig: drawing_ticrcu}) were deposited with identical deposition duration and, within the frame of the IBSD reproducibility, have equal mass depositions. The same is true for the separate titanium-chromium bilayers sample and the bilayers in both multilayer samples.

\begin{figure}
\centering
\includegraphics[keepaspectratio, width=\linewidth]{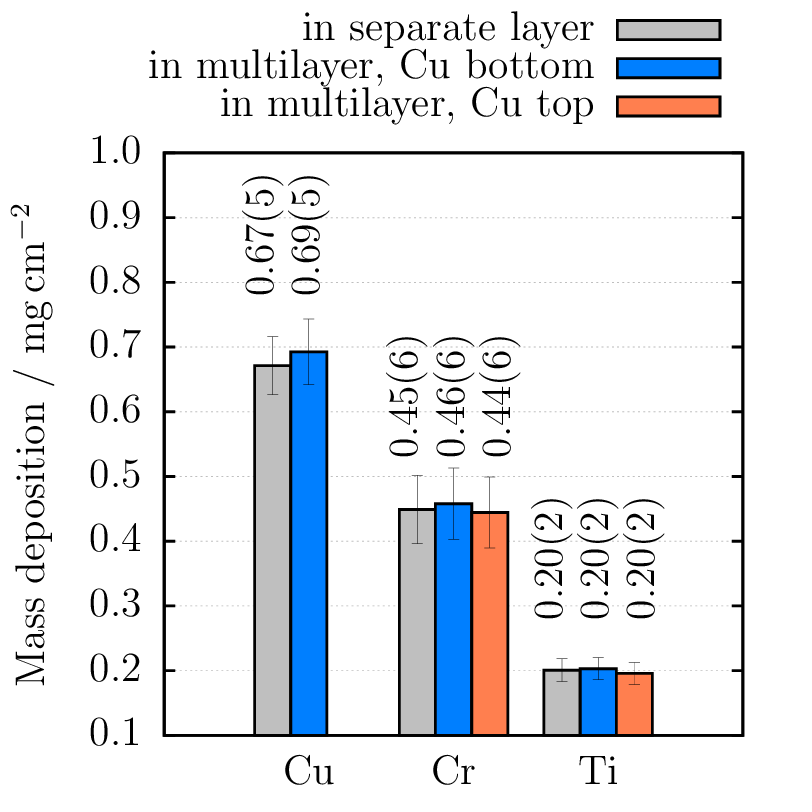}
\caption{Mass depositions of the samples shown in figure~\ref{fig: schematics}, determined with reference-free XRF analysis. Results are shown in comparison between separate layers and layers in a multilayer sample, which were produced with equal deposition durations.}
\label{fig: multiref_result}
\end{figure}

The homogeneity of the samples was verified by XRF measurements at different sample positions. The determined deviations are in the order of the counting statistical uncertainty of the measurement and therefore deemed satisfactory. A quantitative, reference-free XRF analysis was performed with a monochromatic photon energy of \SI{9.8}{\keV}. The results are shown in figure~\ref{fig: multiref_result}. The determined mass deposition and its uncertainty is given for the three constituents copper, chromium, and titanium for the different sample types, i.e., for the separate layers and the two multilayers with different layer sequence (figure \ref{fig: schematics}). Two of the copper layers are approximately $d_{\text{Cu}}\approx\SI{0.76}{\um}$ thick. The third copper layer was omitted here, because it was produced with a similar but not identical deposition duration and therefore cannot directly be compared to the other two values. The thickness of the titanium-chromium alloy layers is about $d_{\text{Ti-Cr}}\approx\SI{1}{\um}$.

Independent validation is given by the comparison of elements in the samples, which were deposited with equal deposition duration: The results of separate and multilayer systems are nearly equal, with deviations significantly lower than the calculated uncertainties, as shown in figure~\ref{fig: multiref_result}. The maximum relative deviation between separate and multilayer samples is given by about \SI{3}{\percent} for the copper mass deposition. Chromium and titanium show even lower discrepancies. While the contribution of secondary fluorescence is significant for both titanium (about \SI{20}{\percent}) and chromium (about \SI{10}{\percent}) in the multilayer samples, only titanium has a secondary enhancement in the separate Ti-Cr alloy (about \SI{10}{\percent}). The agreement between quantitative results of samples with very different absorption paths and secondary fluorescence contributions shows the appropriate consideration of these effects, further validating the secondary fluorescence model.

\section{Angular dependence of secondary fluorescence intensity}

\begin{figure}
\centering
\includegraphics[keepaspectratio]{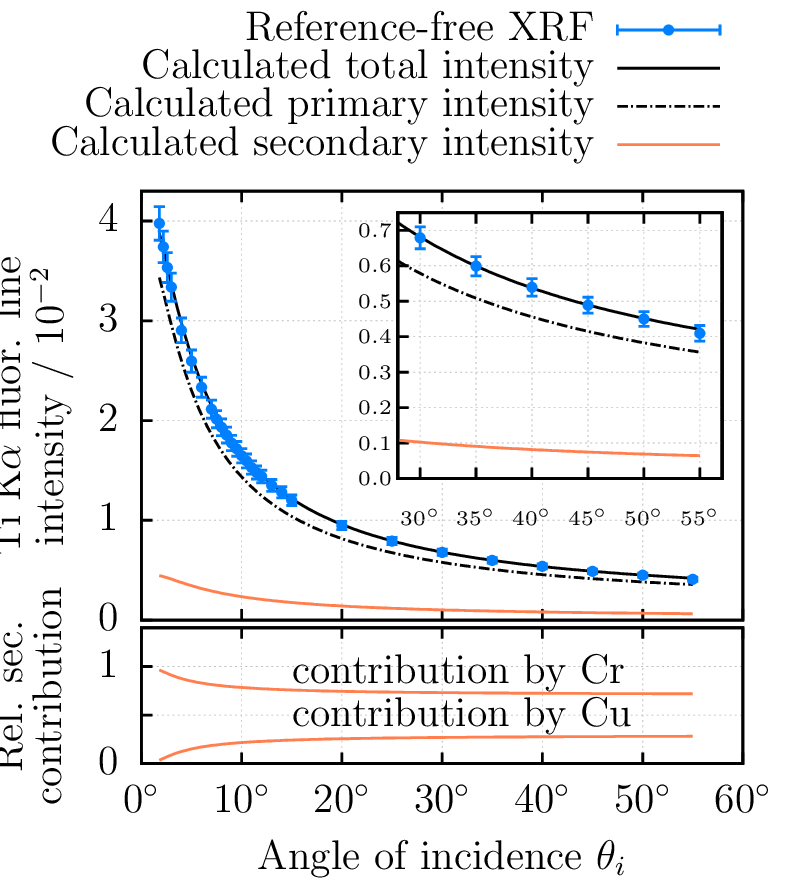}
\caption{Comparison of measured and calculated angular dependency of the titanium K$\alpha$ fluorescence line intensity per incident photon, shown at the top. The calculated relative contributions for chromium and copper to the secondary fluorescence are shown at the bottom.}
\label{fig: angulardist}
\end{figure}

A sample equivalent to the multilayer sample with copper at the bottom (figure \ref{subfig: drawing_ticrcu}) was analysed with reference-free XRF experiments by employing multiple different incident angles $\theta_i$ and a monochromatic excitation energy of \SI{9.8}{\keV}. The incident angle~$\theta_i$ was varied from \SI{1.8}{\degree} to \SI{55}{\degree}, while the angle of detection was always set to $\theta_d=\SI{90}{\degree}-\theta_i$, i.e., in the range of \SI{88.2}{\degree} to \SI{35}{\degree}. Both angles are measured in respect to the sample surface. The mass depositions of the sample layers are determined from the measurement at $\theta_i=\theta_d=\SI{45}{\degree}$. The theoretical angular distribution of the fluorescence intensity according to the Sherman equation \eqref{eq: intensity}, based on these determined mass depositions, can then be compared with the measured angular distribution. This is shown in figure~\ref{fig: angulardist}, for the titanium K$\alpha$ fluorescence line intensity. Furthermore, the predicted primary and secondary fluorescence contributions are given. The total Ti K$\alpha$ fluorescence intensity varies nearly by one order of magnitude over the considered angular range, with the highest slope for low angles of incidence $\theta_i<\SI{20}{\degree}$. The direct comparison of measurements and calculations are in good agreement for the entire angular range, when taking the determined uncertainties into account. This comparison allows for a validation of the secondary fluorescence model with decidedly little assumptions. In essence, these are only the usual prerequisites~\cite{sherman} of the Sherman equation \eqref{eq: intensity}, i.e., homogeneity and smoothness of the sample, which are readily verified. Over the wide angular range the effective information depth differs substantially, since the angle variation changes the X-ray absorption paths~\cite{Ebel1971,Boer1989,fiorini}. Likewise, the relative secondary fluorescence contribution changes from about \SI{10}{\percent} to approximately \SI{14}{\percent} for the titanium K$\alpha$ line. For low incident angles this secondary fluorescence is almost solely produced by chromium in the alloy layer due to intralayer excitation, because nearly no excitation radiation reaches the bottom copper layer. This can be seen from the relative contributions to the secondary fluorescence due to chromium and copper, shown in the bottom figure~\ref{fig: angulardist}. For high incident angles the relative contributions are about \SI{28}{\percent} for copper and \SI{72}{\percent} for chromium.

\section{Selective excitation to disable secondary fluorescence}

Tuning the monochromatic energy of the incident radiation allows for a selective excitation of the elements in the sample~\cite{Iida1985}. This selective excitation can be used to physically disable the secondary fluorescence effect, without moving the sample. One of the multilayer samples, with the copper layer at the top (figure \ref{subfig: drawing_cuticr}), was excited with monochromatic radiation with an energy of $E_0=\SI{8.92}{\keV}<\text{Cu-K}_{abs}$, which is below the binding energy of an electron from the copper K shell $\text{Cu-K}_{abs}$. In this case, copper K fluorescence is not produced, and secondary fluorescence due to copper K fluorescence is impossible. While chromium K fluorescence will still produce secondary fluorescence for titanium, in principle, this can be disabled by measuring with an excitation energy below the chromium K edge.

The copper layer always contributes to the attenuation of the X-ray radiation, when it is spatially above the titanium-chromium alloy in respect to the incident radiation. To account for this, the copper mass deposition determined above the copper K edge is assumed to be known for the lower excitation energy. The mass depositions of titanium~$\rho{}d_{\text{Ti}}$ and chromium~$\rho{}d_{\text{Cr}}$ are readily determined by measurements utilizing both energies with this preconsideration. Since the sample does not change when tuning the energy of the incident radiation, the quantitative results are expected to be equal for both measurements. Indeed, the ratios of determined mass depositions $\frac{\rho{}d_{\text{Cr}}\left(E_0>\text{Cu-K}_{\text{abs}}\right)}{\rho{}d_{\text{Cr}}\left(E_0<\text{Cu-K}_{\text{abs}}\right)}\approx\num{1.01}$ and $\frac{\rho{}d_{\text{Ti}}\left(E_0>\text{Cu-K}_{\text{abs}}\right)}{\rho{}d_{\text{Ti}}\left(E_0<\text{Cu-K}_{\text{abs}}\right)}\approx\num{1.01}$ are close to unity, further validating the secondary fluorescence model. This approach can easily be extended by tuning the energy of the incident radiation across all relevant ionization edges of a sample. The increase in available information is achieved at the cost of an increase in experimental effort. Furthermore, the applicability of utilizing multiple monochromatic energies is usually limited to monochromator beamlines at synchrotron radiation facilities.

\section{Conclusion}

We have demonstrated multiple methodologies for validating the commonly used secondary fluorescence model in XRF. This is achieved by employing reference-free XRF analysis on different layer systems, including artificially created thin alloy layers, all showing a significant contribution of secondary fluorescence between \SI{10}{\percent} to \SI{30}{\percent}. When varying the excitation conditions, e.g., by altering the monochromatic energy of the incidence radiation, the absorption paths inside the sample change substantially and enhancement channels due to secondary fluorescence can be explicitly enabled or disabled. An approach to design and reliably produce alloy layers for the purpose of XRF was described. Since for thin alloyed material no established reference samples exist, this provides the opportunity to qualify appropriate calibration samples, e.g., with reference-free XRF analysis. For the determination of uncertainties, special care was taken to include all contributions of experimental parameters and FPs. The total uncertainties of the quantitative results are in the order of about \SI{10}{\percent}, which mostly result from the uncertainties of the employed FPs. To determine FPs with lower uncertainties, dedicated experiments or calculations are required~\cite{iifpRoadmap}.

\begin{acknowledgement}
The authors would like to thank Jan Weser for support during the experiments. This research was performed within the EMPIR project Aeromet. The financial support of the EMPIR program is gratefully acknowledged. It is jointly funded by the European Metrology Programme for Innovation and Research (EMPIR) and participating countries within the European Association of National Metrology Institutes (EURAMET) and the European Union.
\end{acknowledgement}

\bibliography{references}

\clearpage
\onecolumn
\thispagestyle{empty}
\section*{Graphical TOC Entry}
\begin{center}
\includegraphics[keepaspectratio, width=3.25in, height=1.75in]{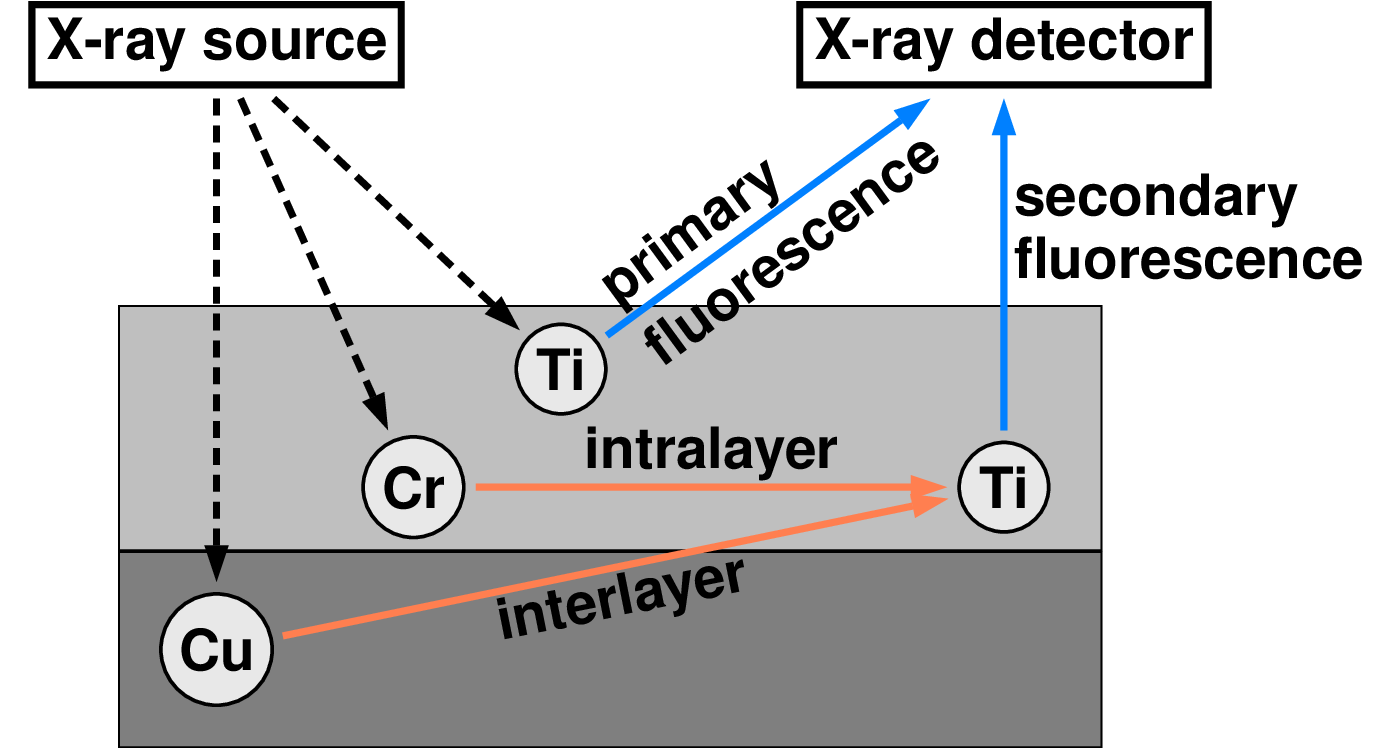}
\end{center}

\end{document}